\documentclass[
 twocolumn,
 superscriptaddress,
 preprintnumbers,
 nofootinbib,
 amsmath,amssymb,
 aps,
 prl,
]{revtex4}
\usepackage{psfrag,slashed,cancel,array,graphicx,hyperref}
\usepackage{subfigure}
\usepackage[utf8]{inputenc}
\usepackage{mathtools}
\usepackage{slashed}
\usepackage{mathrsfs}
\usepackage{multirow}
\usepackage{tikz}
\usepackage[compat=1.1.0]{tikz-feynman}
\usepackage{braket}
\usepackage{float}
\usepackage{booktabs} 
\usepackage[normalem]{ulem}
\hypersetup{pdftitle={},pdfcreator={},linkcolor=[rgb]{0.15,0.35,0.75},colorlinks=true,citecolor=[rgb]{0.675,0,0.2},urlcolor=[rgb]{0.15,0.35,0.65}}

\usepackage{comment}
\usepackage{color} 
\usepackage{graphics}

\newcommand{\asr}{\left(\frac{\alpha_s}{4\pi}\right)}

\newcommand{\oaas}{${\cal O}(\alpha_s\alpha)$\,} 
\newcommand{\oaasas}{${\cal O}(\alpha_s^2\alpha)$\,} 

\newcommand{\sigNLO}{\sigma^{\rm \scriptscriptstyle NLO}}

\def\asr{\bigg( \frac{\alpha_s}{4 \pi} \bigg)}
\def\aer{\bigg( \frac{\alpha}{4 \pi} \bigg)}

\def\mur{\mu_{\scriptscriptstyle R}}
\def\muf{\mu_{\scriptscriptstyle F}}

\def\renM{{\cal M}}

\def\cD{{\cal D}}
\def\lifourH{{\rm Li}_4{\left({\textstyle \frac{1}{2}}\right)}}

\allowdisplaybreaks

\begin{document}

\tikzset{
photon/.style={decorate, draw=black,
    decoration={coil,aspect=0,segment length=3pt,amplitude=1pt}},    
Zboson/.style={decorate, draw=blue,
    decoration={snake,aspect=0,segment length=3pt,amplitude=1pt}},   
Wboson/.style={decorate, draw=red,
    decoration={snake,aspect=0,segment length=3pt,amplitude=1pt}},   
gluon/.style={decorate, draw=black,
    decoration={coil,aspect=0.5,segment length=3pt,amplitude=1pt}},
fermion/.style={draw=black,
      postaction={decorate},decoration={markings,mark=at position .55
        with {\arrow[draw=black]{>}}}},   
scalar/.style={draw=black,
      postaction={decorate},decoration={}},           
vector/.style={decorate, decoration={snake}, draw} 
}


\title{${\mathcal{O}}(\alpha_s^2\alpha)$ threshold corrections to inclusive $Z$-boson production}

\author{Tanmoy Pati}
\email{tanmoy.pati@niser.ac.in}
\affiliation{School of Physical Sciences, National Institute of Science Education and Research, 752050 Jatni, India}
\affiliation{Homi Bhabha National Institute, Training School Complex, Anushakti Nagar, Mumbai 400094, India}

\author{Narayan Rana}
\email{narayan.rana@niser.ac.in}
\affiliation{School of Physical Sciences, National Institute of Science Education and Research, 752050 Jatni, India}
\affiliation{Homi Bhabha National Institute, Training School Complex, Anushakti Nagar, Mumbai 400094, India}

\author{Avick Saha}
\email{avick.saha@niser.ac.in}
\affiliation{School of Physical Sciences, National Institute of Science Education and Research, 752050 Jatni, India}
\affiliation{Homi Bhabha National Institute, Training School Complex, Anushakti Nagar, Mumbai 400094, India}

\date{\today}

\begin{abstract}
To match the experimental precision targeted by future collider programs, theoretical predictions for Drell-Yan production must incorporate higher-order mixed Quantum Chromodynamics and electroweak corrections. In this work, we present for the first time the ${\mathcal{O}}(\alpha_s^2\alpha)$ corrections to the inclusive cross-section for $Z$-boson production in the threshold limit. 
By systematically evaluating the missing singlet contributions,
we obtain the complete set of ${\mathcal{O}}(\alpha_s^2\alpha)$ quark form factors,
which we use to construct the soft-virtual cross-section by exploiting renormalization group invariance, the K+G evolution equation, and universal soft-collinear factorization. Our results capture the dominant threshold logarithmic enhancements, improve the theoretical scale uncertainties, significantly reduce electroweak input scheme dependence, and 
establish a crucial step toward complete ${\mathcal{O}}(\alpha_s^2\alpha)$ corrections indispensable for future precision programs.
\end{abstract}


\pacs{}

\maketitle


\section{Introduction}
The Drell-Yan (DY) process~\cite{Drell:1970wh}, production of a lepton pair in hadronic collisions, plays a fundamental role in probing Quantum Chromodynamics (QCD) and Electroweak (EW) physics,
through the precision determination of 
the proton parton distribution functions (PDFs)~\cite{Amoroso:2022eow}, 
the strong coupling constant $\alpha_s$~\cite{Ball:2018iqk,ATLAS:2023lhg}
and the Standard Model (SM) parameters, such as the $W$ boson mass~\cite{ATLAS:2017rzl,LHCb:2021bjt}, the weak mixing angle~\cite{Aaltonen:2018dxj,LHCb:2015jyu}.
On-shell Z-boson production approximates resonant neutral-current (NC) DY.
Due to its large cross-section and clean leptonic decay modes, this process serves as a vital benchmark at hadron colliders.
Higher-order predictions for the production of a vector boson have advanced through 
the next-to-leading-order~(NLO)~\cite{Altarelli:1979ub},
next-to-next-to-leading-order~(NNLO)~\cite{Matsuura:1987wt,Matsuura:1988nd,Hamberg:1990np,Matsuura:1990ba,vanNeerven:1991gh,Harlander:2002wh,Moch:2005ky,deFlorian:2012za,Ahmed:2016qhu,Anastasiou:2003yy,Anastasiou:2003ds,Melnikov:2006kv,Catani:2009sm,Catani:2010en}, 
and next-to-next-to-next-to-leading-order (N$^3$LO)~\cite{Ahmed:2014cla,Catani:2014uta,Li:2014afw,Duhr:2020seh,Chen:2021vtu,Camarda:2021ict,Chen:2022cgv,Neumann:2022lft,Campbell:2023lcy} QCD corrections,
reaching sub-percent accuracy in theoretical predictions and systematically reducing scale uncertainties.

The precision targets in present and future colliders have driven the theoretical 
studies to go beyond the high-precision QCD framework to include EW effects. 
The EW corrections not only make the predictions more accurate, 
but also are essential to reduce the scheme dependence arising from the choice of different input parameters.
While NLO EW corrections to NC DY~\cite{Baur:2001ze,Zykunov:2005tc,CarloniCalame:2007cd,Arbuzov:2007db,Dittmaier:2009cr}
and charged current (CC) DY~\cite{Dittmaier:2001ay,Baur:2004ig,Zykunov:2006yb,Arbuzov:2005dd,CarloniCalame:2006zq}
were established over two decades ago, 
recent years have seen significant progress toward mixed QCD-EW $\mathcal{O}(\alpha_s \alpha)$ corrections,
where $\alpha$ is the electromagnetic coupling constant.
Initial works focused on mixed QCD-QED corrections to inclusive on-shell $Z$-boson production~\cite{deFlorian:2018wcj,Cieri:2020ikq,Delto:2019ewv}, 
followed by complete mixed QCD-EW corrections for on-shell $W$- and $Z$-boson production~\cite{Bonciani:2016wya,Bonciani:2019nuy,Bonciani:2020tvf,Buccioni:2020cfi,Behring:2020cqi,Bonciani:2021iis}. 
To extend predictions beyond the on-shell limit, the pole approximation~\cite{Denner:2019vbn} was employed to capture dominant resonant contribution effects~\cite{Dittmaier:2014qza,Dittmaier:2015rxo,Dittmaier:2024row}. 
Most recently, complete mixed QCD-EW corrections for both NC and CC DY processes across the full off-shell phase space have been achieved in 
refs.~\cite{Bonciani:2016ypc,Heller:2019gkq,Heller:2020owb,Dittmaier:2020vra,Buonocore:2021rxx,Bonciani:2021zzf,Buccioni:2022kgy,Armadillo:2022bgm}.
The numerical impact of complete $\mathcal{O}(\alpha_s \alpha)$ mixed QCD-EW corrections proved to be larger than anticipated across various kinematic regions, pushing the efforts toward 
complete NNLO EW corrections~\cite{Armadillo:2025mfx,Freitas:2025vax} and 
three-loop \oaasas mixed QCD-EW corrections~\cite{Pati:2025ivg,Pati:2025xht}. 
Both sets of corrections are formally required to match the high experimental 
precision demanded by the future collider programs.  
Furthermore, $\mathcal{O}(\alpha_s \alpha)$ predictions contain theoretical uncertainties stemming from 
renormalization and factorization scale variations and the choice of EW input parameters. 
Incorporating \oaasas corrections is therefore essential to stabilize the perturbative expansion, 
reduce the residual uncertainties, and ensure theoretical predictions keep pace with experimental precision.

In this work, we address this requirement by presenting, for the first time, \oaasas corrections to the inclusive production 
cross-section of the $Z$-boson in the threshold limit. We systematically evaluate the missing singlet contributions 
to the three-loop \oaasas form factors. Further, we exploit the K+G equation, renormalization group (RG) invariance, and
the universal structure of QCD factorization to obtain the Soft-Virtual (SV) threshold cross-section for inclusive $Z$-boson production at \oaasas.
Our calculation is performed within the SV framework developed in refs.~\cite{Ravindran:2005vv,Ravindran:2006cg,Ahmed:2014era,Ahmed:2014cha,Ahmed:2015qda,Ahmed:2016otz,Ahmed:2020nci,Goyal:2025bzf,AH:2019xds,AH:2019pyp}, 
which systematically accounts for soft and collinear radiation. 
While the SV approximation does not capture the full hard-real emission spectrum, 
these results capture the dominant logarithmic threshold enhancements,
improve the scale uncertainties, significantly reduce the EW input scheme dependence,  
and marks the first step toward complete ${\mathcal{O}}(\alpha_s^2\alpha)$ corrections essential for future colliders.

The remainder of this paper is organized as follows: In the next section, we outline the theoretical framework. We then present the numerical results and discuss their phenomenological impact. Finally, we provide our concluding remarks.

\section{Theoretical framework}
The $Z$-boson inclusive production cross-section $\sigma_{\text{tot}}$ in hadron colliders can be expressed as 
\begin{align}
\sigma_{\text{tot}} (\tau) = \sum_{a,b} \int {\rm d}x_1 {\rm d}x_2 \, {f}_a (x_1) {f}_b (x_2) \, \sigma_{ab} (z) \, ,
\label{eq:sigmatot-bare}
\end{align}
where the sum runs over quarks ($q$), anti-quarks ($\bar{q}$), gluons ($g$), and photons ($\gamma$).
The dimensionless scaling variables are defined as 
$\tau = \frac{m_Z^2}{S}$ and $z = \frac{m_Z^2}{\hat{s}}$, 
where $m_Z$ is the mass of the $Z$-boson, with $S$ and $\hat{s}$ being the hadronic and partonic center-of-mass energies squared, respectively. $S$ and $\hat{s}$ are related through the Bjorken momentum fractions $x_1, x_2$ by $\hat{s} = x_1 x_2 S$.
The finite partonic cross-section $\sigma_{ab}$ is convoluted with the renormalized PDFs. 
In the SM, $\sigma_{ab}$ can be factorized into a Born prefactor $\sigma^{(0)}$ and 
a dimensionless coefficient function (CF) $\Delta_{ab}(z)$, as $\sigma_{ab}(z) = \sigma^{(0)} \Delta_{ab}(z)$. 
Expansion in $\alpha_s$ and $\alpha$ yields:
\begin{equation} 
\Delta_{ab} (z) = \sum_{m,n=0}^{\infty}\alpha_s^m \alpha^n ~ \Delta_{ab}^{(m,n)} (z) \, ,
\label{eq:deltaexp}
\end{equation}
where $\Delta_{ab}^{(m,n)}$ denotes the $\mathcal{O}(\alpha_s^m\alpha^n)$ correction to $\Delta_{ab}(z)$.

For a given initial state, $\Delta_{ab}(z)$ receives contributions from processes with different 
final-state multiplicities due to real parton emissions. The logarithmic structure that arises in the soft-collinear 
limit exhibits universal behavior, allowing for all-order resummation and thus motivating a systematic decomposition 
of $\Delta_{ab}(z)$. In the threshold limit $z \to 1$, $\Delta_{ab}(z)$ can be decomposed into a 
SV part and a regular part as follows:\begin{equation}
\label{eq:deltaSV}
\Delta_{ab} (z) = \delta_{\bar{a}b} \, \Delta_{a \bar{a}}^{\text{sv}}(z) + \Delta_{ab}^{\text{reg}}(z) \, ,
\end{equation}
where $\delta_{\bar{a}b}$ ensures the SV contributions from diagonal partonic channels.
The SV term $\Delta_{a\bar{a}}^{\text{sv}}(z)$ contains singular distributions, known as threshold logarithms,
expressed in terms of plus distributions
as ${\mathcal{D}}_n = \big( \frac{\ln^n (1-z)}{1-z} \big)_{\scriptscriptstyle +}$.
The plus distribution is defined by 
\begin{equation}
 \int_{0}^1 dx f(x) \left( \frac{1}{x} \right)_+ = \int_0^1 dx \left( \frac{f(x)-f(0)}{x} \right) \,.
\end{equation}
The regular part, $\Delta_{ab}^{\text{reg}}(z)$, contains terms that remain finite in the threshold limit $z \rightarrow 1$.

Governed by Sudakov resummation and RG equations, the SV CF 
$\Delta_{q\bar{q}}^{\text{sv}}(z) \equiv \Delta_{q}^{\text{sv}}(z)$
exponentiates into a finite distribution $\Psi_q$ in the 
threshold limit~\cite{Ravindran:2005vv,Ravindran:2006cg}:
\begin{equation}
\label{sigma}
 \Delta^{\text{sv}}_{q} (z, q^2, \mur^2, \muf^2) = 
{\cal C} \exp \left( \Psi_q (z, q^2, \mur^2, \muf^2, \epsilon)  \right) \big|_{\epsilon = 0}
\end{equation}
where the symbol ${\cal C}$ denotes Mellin convolution, given by
\begin{equation}
\label{conv}
 {\cal C} e^{f(z)} = \delta(1-z) + \frac{1}{1!} f(z) + \frac{1}{2!} f(z) \otimes f(z) + \ldots 
\end{equation}
Here, $f(z)$ comprises singular distributions containing $\delta(1-z) \equiv \delta_z$ 
and ${\cal D}_n$. All regular terms in $z$ are systematically dropped in intermediate steps.
The exponent $\Psi_q$ factorizes into hard, soft, and collinear components, as \begin{align}
\label{eq:psi}
\Psi_q = \ln | F_q |^2 \delta(1-z) + 2 \, \Phi_q - 2 \, {\cal C} \ln \Gamma_{qq} \, .
\end{align}
where $F_q$ denotes the ultraviolet (UV) renormalized virtual matrix element squared, 
$\Phi_q$ is the soft distribution function capturing soft-gluon and soft-photon emissions, 
and $\Gamma_{qq}$ represents the diagonal space-like Altarelli-Parisi splitting kernel. 
The universal behavior of $\Phi_q$ and $\Gamma_{qq}$ in mixed QCD-QED framework has been detailed extensively in refs.~\cite{AH:2019xds,AH:2019pyp}.
In extending this formalism from mixed QCD-QED to mixed QCD-EW theory, 
the structural forms of $\Phi_q$ and $\Gamma_{qq}$ remain unchanged, 
as massive gauge bosons do not generate infrared (IR) singularities. 
Consequently, weak-interaction effects contribute exclusively to the 
process-dependent virtual contributions $F_q$.
In the following, we outline the formal definition of $F_q$ and 
detail the calculation of the missing contributions at \oaasas.

\subsection{The Pure Virtual Contributions}
$F_q$ admits a double series expansion in $\alpha_s$ and $\alpha$ as
\begin{equation}
 F_q = \sum_{m,n} \asr^m \aer^n \frac{\langle \renM^{(0,0)} | \renM^{(m,n)} \rangle}{\langle \renM^{(0)} | \renM^{(0)} \rangle}.
\end{equation}
$| \renM^{(m,n)} \rangle$ is the ${\mathcal{O}}(\alpha_s^m \alpha^n)$ corrections to
the UV renormalized amplitude. 
$\langle \renM^{(0)} | \renM^{(0)} \rangle$ is defined by
$\langle \renM^{(0,0)} | \renM^{(0,0)} \rangle = (v_q^2 + a_q^2) \langle \renM^{(0)} | \renM^{(0)} \rangle$
where,
$v_q = ( {I_3^q/2} - s_w^2 e_q ) \,, a_q = {I_3^q}/2$.
$I_3^q$ is the third component of the quark isospin,
$e_q$ is the electric charge of the initial state quark
and 
$s_w$ is the sine of weak mixing angle.

Radiative corrections to the axial-vector current introduce an additional challenge 
due to the presence of $\gamma_5$, which is inherently a four-dimensional object. 
The topology of the fermion lines containing $\gamma_5$ distinguishes two classes of diagrams: 
non-singlet diagrams, where an even number of $\gamma_5$ matrices resides on each fermion line, 
and singlet diagrams, where an odd number of $\gamma_5$ matrices appears on each fermion line. 
For non-singlet diagrams, one can consistently employ 
the Naive Dimensional Regularization (NDR) scheme, which preserves the anti-commutativity of 
$\gamma_5$. For singlet diagrams, however, the NDR scheme is inadequate. 
Over the years, several schemes have been developed to resolve this issue 
by relaxing either anti-commutativity or the standard cyclicity of the trace.
Among them, the scheme proposed in refs.~\cite{tHooft:1972tcz,Breitenlohner:1977hr,Larin:1991tj,Larin:1993tq}, 
is widely adopted. In this scheme, $\gamma_5$ is defined as
\begin{equation}
\gamma_\mu \gamma_5 
 = \frac{i}{3!} \epsilon_{\mu\nu\rho\sigma} \gamma^\nu \gamma^\rho \gamma^\sigma.  
\label{eq:g5def}
\end{equation}
As it abandons anti-commutativity, the corresponding axial Ward identities are violated. 
To restore these identities, additional renormalization constants must be introduced order by order 
in perturbation theory. 
These renormalization constants have been calculated up to 
four loops in pure QCD in refs.~\cite{Larin:1991tj, Larin:1993tq, Ahmed:2015qpa, Ahmed:2021spj, Chen:2021gxv, Chen:2022lun}, 
and at \oaasas for mixed QCD-QED in ref.~\cite{Pati:2026upt}. 
As introducing a mass to the U(1) gauge boson does not alter the UV structure, 
these same constants remain applicable to \oaasas mixed QCD-Weak corrections 
for topologies containing at most two $\gamma_5$ matrices.
In principle, \oaasas mixed QCD-Weak corrections also admit singlet contributions 
where one fermion line contains three $\gamma_5$ matrices and another contains one. 
However, much like the anomalous contributions in pure QCD, these additional terms 
cancel pairwise within a massless quark flavor doublet. Hence, for even number of quark flavors, 
the non-vanishing singlet contributions originate exclusively from topologies involving 
only two $\gamma_5$ matrices. The constants derived in ref.~\cite{Pati:2026upt} 
are therefore fully sufficient to construct UV-finite singlet \oaasas form factors.

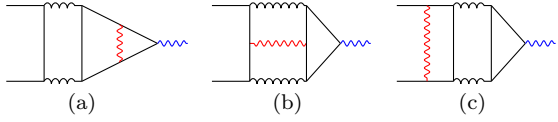
\begin{figure}[h]
\centering
\begin{tikzpicture}[scale=0.5]
 \draw[scalar]  (-4,1) -- (-3,1);
 \draw[gluon]  (-3,1) -- (-2,1);
 \draw[scalar]  (-2,1) -- (0,0);
 \draw[scalar]  (-4,-1) -- (-3,-1);
 \draw[gluon]  (-3,-1) -- (-2,-1);
 \draw[scalar]  (-2,-1) -- (0,0);
 \draw[Zboson] (0,0) -- (0.8,0); 
 \draw[Wboson]  (-1,0.5) -- (-1,-0.5);
 \draw[scalar]  (-2,-1) -- (-2,1); 
 \draw[scalar]  (-3,-1) -- (-3,1); 
 \node at (0.9,0) {$~$};
 \node at (-2,-1.6) {\footnotesize (a)};
 
\end{tikzpicture}\hspace{0.1cm}\begin{tikzpicture}[scale=0.5]
 \draw[scalar]  (-4,1) -- (-3,1);
 \draw[gluon]  (-3,1) -- (-1.5,1);
 \draw[scalar]  (-1.5,1) -- (-0.6,0);
 \draw[scalar]  (-4,-1) -- (-3,-1);
 \draw[gluon]  (-3,-1) -- (-1.5,-1);
 \draw[scalar]  (-1.5,-1) -- (-0.6,0);
 \draw[Zboson] (-0.6,0) -- (0.2,0); 
 \draw[Wboson]  (-1.5,0) -- (-3,0);
 \draw[scalar]  (-1.5,-1) -- (-1.5,1); 
 \draw[scalar]  (-3,-1) -- (-3,1); 
 \node at (0.35,0) {$~$};
 \node at (-2,-1.6) {\footnotesize (b)};
 
\end{tikzpicture}\hspace{0.1cm}\begin{tikzpicture}[scale=0.5]
 \draw[scalar]  (-4,1) -- (-2.5,1);
 \draw[gluon]  (-2.5,1) -- (-1.5,1);
 \draw[scalar]  (-1.5,1) -- (-0.6,0);
 \draw[scalar]  (-4,-1) -- (-2.5,-1);
 \draw[gluon]  (-2.5,-1) -- (-1.5,-1);
 \draw[scalar]  (-1.5,-1) -- (-0.6,0);
 \draw[Zboson] (-0.6,0) -- (0.2,0); 
 \draw[Wboson]  (-3.2,1) -- (-3.2,-1);
 \draw[scalar]  (-1.5,-1) -- (-1.5,1); 
 \draw[scalar]  (-2.5,-1) -- (-2.5,1); 
 \node at (-2,-1.6) {\footnotesize (c)};
\end{tikzpicture}
\caption{Schematic representations of the singlet diagrams.}
\label{fig:feyndia}
\end{figure}

Virtual corrections are established at ${\mathcal{O}}(\alpha_s^3)$~\cite{Moch:2005tm,Baikov:2009bg,Gehrmann:2010ue,Ahmed:2016qjf}, 
at \oaas~\cite{Bonciani:2021iis}, and \oaasas non-singlet~\cite{Pati:2025xht}. 
In this paper, we compute the remaining \oaasas contributions to $F_q$.
Schematic representations of the relevant Feynman diagrams are shown in Fig.~\ref{fig:feyndia}. 
These matrix elements can be defined as 
\begin{align}
\frac{\langle \mathcal{M}^{(0,0)}|\mathcal{M}^{(2,1)}_{\text{s}} \rangle}{\langle \mathcal{M}^{(0)}|\mathcal{M}^{(0)} \rangle}
   &=  \frac{a_u \sum a_f e_f^2}{c_w^2 s_w^2} F_{\gamma,\text{s}}^{(2,1)} 
    + \frac{a_u \sum v_f^2 a_f}{c_w^4 s_w^4} F_{{\scriptscriptstyle Z},\text{s}}^{(2,1)} 
\nonumber\\
   &+ \frac{a_u \sum v_f}{8 c_w^2 s_w^4} F_{{\scriptscriptstyle W},\text{s}}^{(2,1)} .
\end{align}
$\sum$ implies the sum over even number of quark flavors. 
The subscript $u$ represents the incoming quark flavor,
and $c_w^2 = 1-s_w^2$.

We employ a standard multi-loop computational framework to obtain these three-loop amplitudes: Feynman diagrams 
are generated using \texttt{QGRAF}~\cite{Nogueira:1991ex}, algebraic tensor manipulations are performed with 
in-house \texttt{FORM}~\cite{Vermaseren:2000nd,Ruijl:2017dtg} routines, and integration-by-parts (IBP) reductions 
to master integrals are carried out using \texttt{Kira}~\cite{Maierhofer:2017gsa,Klappert:2020nbg} and \texttt{LiteRed}~\cite{Lee:2012cn}, 
followed by UV renormalization of the quark wave function and $\alpha_s$. 
Although the definition of singlet diagrams primarily concerns the axial-vector current, diagrams as shown in
Fig.~\ref{fig:feyndia}(b), yield non-zero contributions for the 
external vector current. For the purely vectorial contributions and those arising from diagrams with an even 
number of $\gamma_5$ matrices on each fermion line, both of which can be consistently treated within the NDR scheme, 
we define the matrix element as
\begin{equation}
\frac{\langle \mathcal{M}^{(0,0)}|\mathcal{M}^{(2,1)}_{\text{Box}}\rangle}{\langle \mathcal{M}^{(0)}|\mathcal{M}^{(0)}\rangle} 
=\frac{(v_u^2+a_u^2)\sum_f (v_f^2+a_f^2)}{c_w^4 s_w^4} F_{Z,\text{Box}}^{(2,1)}.
\end{equation}
Note that remaining contributions from these diagrams involving an odd number of $\gamma_5$ matrices on each fermion line, are absorbed into the anomalous terms.

\section{Results}
In this section, we present our analytical and numerical results. 
All numerical evaluations are performed using the following input parameters:
$m_Z$ = 91.153 GeV,
$m_W$ = 80.358 GeV,
the top quark mass $m_t$ = 172.5 GeV,
the Fermi constant $G_\mu=1.166 378 \times10^{-5}$ GeV$^{-2}$,
and the fine structure constant $\alpha(0)= 7.29735\times10^{-3}$. 
We consider 5 active flavors in proton, employing the
\texttt{NNPDF31\_nnlo\_as\_0118\_luxqed} PDF set.

We begin by presenting the vectorial contributions from singlet topologies, specifically the internal fermion-loop box diagrams (Fig.~\ref{fig:feyndia}-b). 
At the $Z$-boson resonance threshold, the value of $F_{Z,\text{Box}}^{(2,1)}$ is given by
\begin{equation}
 F_{Z,\text{Box}}^{(2,1)} = C_F ( -49.15096 + i ~ 18.12604 ).
\end{equation}

Further we examine the axial-vector contributions to the singlet diagrams.
The analytical expressions for the finite remainders of all three form factors 
are provided in the Appendix.

We next evaluate the \oaasas inclusive cross-section in the threshold limit. 
The primary result of this work is $\Delta_{q,{\rm sv}}^{(2,1)}$, 
which is decomposed into three distinct terms based on the intermediate gauge boson involved:
\begin{equation}
 \Delta_{q,{\rm sv}}^{(2,1)} = 
  \Delta_{q,{\rm sv}}^{(2,1),\gamma} 
+ \Delta_{q,{\rm sv}}^{(2,1),{\scriptscriptstyle Z}} 
+ \Delta_{q,{\rm sv}}^{(2,1),{\scriptscriptstyle W}} .
\end{equation}
We note that $\Delta_{q,{\rm sv}}^{(2,1),\gamma}$, as reported in ref.~\cite{AH:2019pyp}, omitted the singlet contributions. 
In the Appendix, we provide $\Delta_{q,{\rm sv}}^{(2,1),\gamma}$ analytically, 
$\Delta_{q,{\rm sv}}^{(2,1),{\scriptscriptstyle Z}}$ partially analytically,
and $\Delta_{q,{\rm sv}}^{(2,1),{\scriptscriptstyle W}}$ numerically.

\begin{figure}[H]
    \centering
    \includegraphics[width=0.9\linewidth]{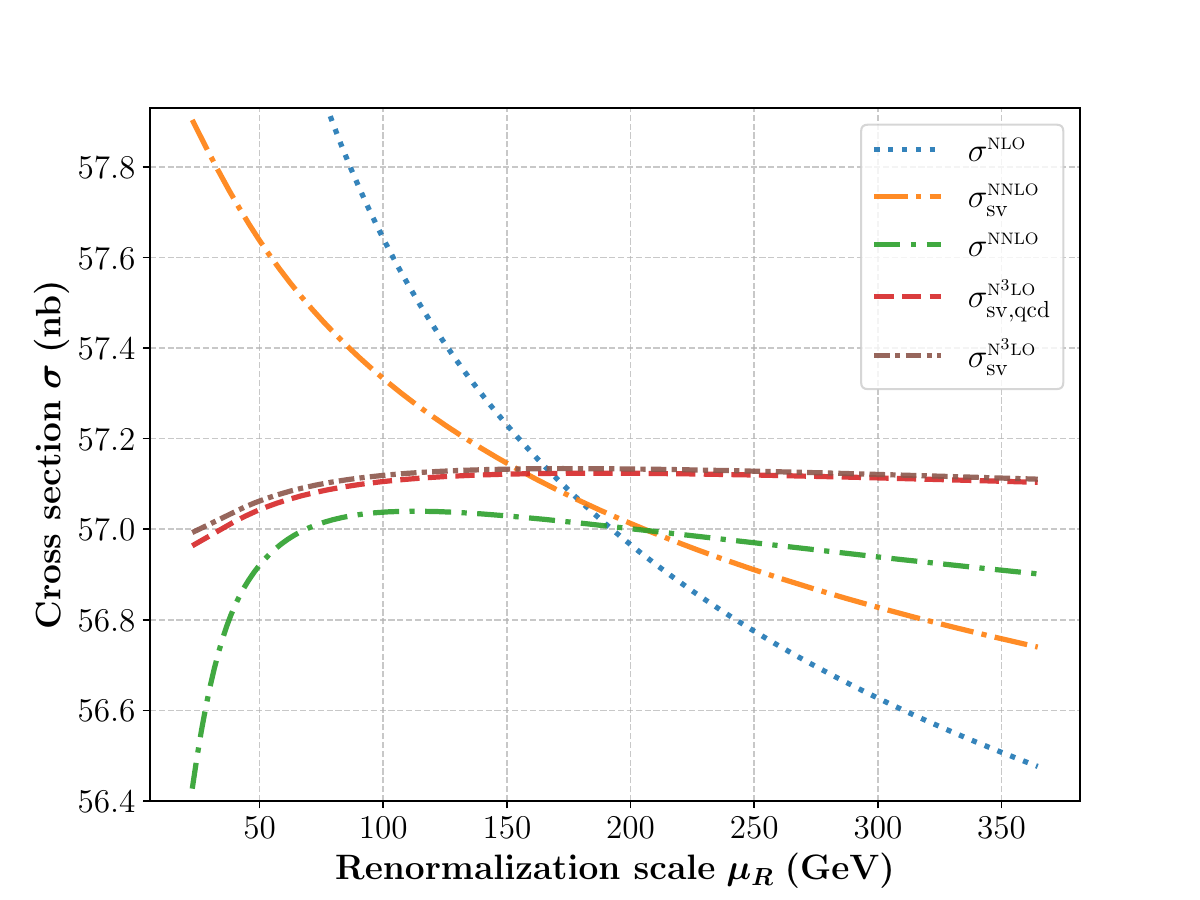}
    \caption{$\mu_R$ variation of the cross-sections at $\sqrt{s}=13$ TeV.}
    \label{fig:muRvar}
\end{figure}

Each individual contribution of $\Delta_{ij}^{(m,n)}$ through convolution with the PDFs
to $\sigma_{\rm tot}$ can be denoted as $\sigma^{(m,n)}$:
\begin{equation}
 \sigma^{(m,n)} = \sum_{i,j} \int {\rm d}x_1 {\rm d}x_2  {f}_i (x_1) {f}_j (x_2) 
 ~ \sigma_{ij}^{(m,n)}.
\end{equation}
where 
$\sigma_{ij}^{(m,n)} = \sigma^{(0)} \alpha_s^m \alpha^n \Delta_{ij}^{(m,n)}$
(no summation over $m,n$).
When $\Delta_{ij}^{(m,n)}$ is restricted to only SV contributions, 
the contribution to $\sigma$ is denoted by `${\rm sv}$' in the subscript e.g $\sigma^{(2,1)}_{{\rm sv}}$.
Further, we define  
$\sigma^{\rm \scriptscriptstyle NLO} = \sigma^{(0,0)} + \sigma^{(1,0)} + \sigma^{(0,1)}$,
$\sigma^{\rm \scriptscriptstyle NNLO} = \sigma^{\rm \scriptscriptstyle NLO} + \sigma^{(1,1)} + \sigma^{(2,0)}$,
$\sigma^{\rm \scriptscriptstyle NNLO}_{\rm sv} = \sigma^{\rm \scriptscriptstyle NLO} + \sigma^{(1,1)}_{\rm sv} + \sigma^{(2,0)}_{\rm sv}$,
$\sigma^{\rm \scriptscriptstyle N^3LO}_{\rm sv} = \sigma^{\rm \scriptscriptstyle NNLO} + \sigma^{(2,1)}_{\rm sv} + \sigma^{(3,0)}_{\rm sv}$,
and
$\sigma^{\rm \scriptscriptstyle N^3LO}_{\rm sv,qcd} = \sigma^{\rm \scriptscriptstyle NNLO} + \sigma^{(3,0)}_{\rm sv}$.
We now present the numerical impact of $\sigma^{(2,1)}_{{\rm sv}}$ on inclusive $Z$-boson production.
Table~\ref{tab:Zcs} presents the hadronic cross-sections across various LHC center-of-mass energies ($\sqrt{s} = 8,13,14,21,28$ TeV) 
and perturbative orders for the central scale choice $\mu_R = \mu_F = m_Z$, evaluated in the $G_\mu$ input parameter scheme ($G_\mu, m_W, m_Z$).
To estimate theoretical scale uncertainties, 
we first vary $\mu_R$ from $m_Z/4$ to $4 m_Z$. 
The resulting upper and lower deviations relative to the central predictions are presented 
as superscripts and subscripts in Table~\ref{tab:Zcs}, respectively.
Overall, the inclusion of \oaasas corrections leads to a noticeable reduction in scale uncertainty.
On the other hand, $\mu_F$ variations yield a mild increase in scale uncertainty.
\begin{widetext}
\begin{center}
\begin{table}[h]
\centering
\setlength{\tabcolsep}{5pt}
\renewcommand{\arraystretch}{1.5}
\begin{tabular}{c | c c c c c c}
\hline
 $\sqrt{s}$ (TeV) & 8 & 13 & 14 & 21 & 28 \\ 
\hline \hline 
 $\sigNLO$ & $33.930^{+1.240}_{-0.801}$ & $57.743^{+1.960}_{-1.267}$ & $62.465^{+2.098}_{-1.356}$ & $95.035^{+3.024}_{-1.955}$ & $126.749^{+3.896}_{-2.518}$ \\ \hline
 $\sigma^{\text{NNLO}}_{\rm sv}$ & $33.723^{+0.355}_{-0.412}$ & $57.375^{+0.529}_{-0.635}$ & $62.065^{+0.561}_{-0.677}$ & $94.410^{+0.773}_{-0.958}$ & $125.904^{+0.969}_{-1.221}$ \\ \hline
 $\sigma^{\text{NNLO}}$& $33.685^{+0.001}_{-0.261}$ & $57.033^{+0.007}_{-0.606}$ & $61.658^{+0.010}_{-0.677}$ & $93.538^{+0.048}_{-1.173}$ & $124.578^{+0.095}_{-1.654}$ \\ \hline
 $\sigma^{\text{N$^3$LO}}_{\text{ sv,qcd}}$ & $33.709^{+0.017}_{-0.053}$ & $57.099^{+0.024}_{-0136}$ & $61.737^{+0.033}_{-0.165}$ & $93.691^{+0.113}_{-0.395}$ & $124.810^{+0.200}_{-0.616}$ \\ \hline
 $\sigma^{\text{N$^3$LO}}_{\text{ sv}}$ & $33.718^{+0.025}_{-0.057}$ & $57.114^{+0.020}_{-0.121}$ & $61.752^{+0.028}_{-0.149}$ & $93.714^{+0.101}_{-0.367}$ & $124.841^{+0.182}_{-0.576}$ \\ \hline
\end{tabular}
\caption{Cross-sections for on-shell $Z$ production, expressed in nanobarns and computed at different LHC energies.}
\label{tab:Zcs}
\end{table}    
\end{center}

\vspace{-0.6cm}

\end{widetext}
Figure~\ref{fig:muRvar} displays the $\mu_R$ dependence of the cross-section 
at $\sqrt{s}= 13$ TeV with $\mu_F=m_Z$. The addition of $\sigma^{(2,1)}_{\rm sv}$ improves 
the $\mu_R$ variation of the cross-section. 
Up to this point, our predictions have been evaluated using the $G_\mu$-scheme. 
Next, we compare the $G_\mu$-scheme with the $\alpha(0)$-scheme, in which $\alpha(0), m_W$ and $m_Z$ serve as the independent 
input parameters. Table~\ref{tab:scheme} illustrates the impact of the \oaasas 
threshold corrections in reducing the relative difference between the two schemes, 
defined as $\delta_s \equiv \delta_{G_\mu-\alpha(0)}$.
\begin{table}[h]
\centering
 \renewcommand{\arraystretch}{1.5}  
 \setlength{\tabcolsep}{3.5pt}    
 \begin{tabular}{|c|c|c|c|c|c|c|}
        \hline
        {Order} & $\sigma^{\text{LO}}$ & $\sigma^{\text{NLO}}$ &   $\sigma^{\text{NNLO}}_{\text{sv}}$ & $\sigma^{\text{NNLO}}$ & $\sigma^{\text{N$^3$LO}}_{\text{sv,qcd}}$ & $\sigma^{\text{N$^3$LO}}_{\text{sv}}$ \\ 
        \hline
        \hline
        {$\delta_{s}\;(\%)$}  & 3.411 & 1.019 & 0.641 & 0.202 & 0.206 & 0.098\\
        \hline
 \end{tabular}
 \caption{Scheme dependence with $\mu_R,\mu_F=m_Z$ at $\sqrt{s}=13\text{ TeV}$.}
 \label{tab:scheme}
\end{table}    
We find, at NNLO, hard-real emissions play a crucial role in reducing the scheme dependence. 
Also, while EW corrections reduce scheme dependence, adding higher-order pure QCD corrections 
can reintroduce larger scheme variations. Finally, incorporating $\sigma^{(2,1)}_{\rm sv}$ 
significantly reduces the scheme difference to below $0.1\%$.
The remarkable agreement between the two scheme predictions confirms 
that EW input scheme dependence is effectively eliminated upon incorporating the \oaasas SV corrections.

\section{Conclusions}
In this work, we have presented, for the first time, the calculation of the \oaasas mixed QCD-EW 
corrections to inclusive $Z$-boson production in the threshold limit. 
We systematically evaluated the missing contributions to the 
three-loop \oaasas quark form factors, considering even number of quark flavors. 
Combining the non-singlet and singlet form factors with universal soft-collinear operator 
and renormalization group evolution equations, we constructed the \oaasas threshold cross-section.
Further, we present the phenomenological impact of the newly derived $\sigma^{(2,1)}_{\rm sv}$ 
corrections across different LHC center-of-mass energies. These corrections contribute at the per-mille level to the total hadronic cross-section, providing an improvement in residual scale uncertainties. 
Next, we highlight the crucial role of \oaasas corrections in reducing EW input parameter 
scheme dependence. While pure QCD higher-order corrections increase the EW input scheme uncertainty, incorporating the $\sigma^{(2,1)}_{\rm sv}$ 
corrections significantly reduces the difference between the two  
schemes to below $0.1\%$. This remarkable convergence demonstrates that mixed three-loop 
corrections are essential to eliminate parameter-choice ambiguities in precision EW observables.
Although the threshold approximation omits hard-real emissions, it captures the dominant 
threshold logarithmic enhancements. The results presented here refine the theoretical predictions 
per mille level for inclusive $Z$-boson production and provide the first step toward complete \oaasas calculations.

\section*{Acknowledgment}
We thank S.~Moch, V.~Ravindran, and A.~Vicini for fruitful discussions. 
N.R. thanks the University of Hamburg for their kind hospitality where 
part of this work was carried out.

\bibliographystyle{apsrev4-1} 
\bibliography{main}

\begin{widetext}


\begin{center}
{\large\bfseries Appendix}
\end{center}

Below, we collect the analytical expressions for the IR finite form factors 
$F_{\gamma,\text{s}}^{(2,1)}$, $F_{{\scriptscriptstyle Z},\text{s}}^{(2,1)}$
and $F_{{\scriptscriptstyle W},\text{s}}^{(2,1)}$
for on-shell $Z$-boson production,
considering even number of quark flavors and $\mu_R^2 = m_Z^2$.
\begin{align}
F_{\gamma,s}^{(2,1)}
    &=C_F T_F\bigg\{ 32 \zeta_2 - i \pi (48) \bigg\}\,.
\\
F_{{\scriptscriptstyle Z},\text{s}}^{(2,1)} &=
 2 C_F T_F \Big\{ 1536 ~ \text{Li}_4 \Big(\frac{1}{2}\Big) + 64 \ln^4 (2) 
 - 314 \zeta_2 - 384 \ln^2 (2) \, \zeta_2 - \frac{2922 \zeta_2^2}{5} + 1350 \zeta_3 
 + i \pi \Big( 408 - 192 \ln (2)  
 \nonumber\\
 & 
 - 66 \zeta_2 + 288 \ln (2) ~ \zeta_2 - 414 \zeta_3 \Big) \Big\} \,.
 \\
 F_{{\scriptscriptstyle W},\text{s}}^{(2,1)} &=
 C_F T_F ~ x^{-2}  \Big\{
-32 P_1 \big(
        2 G_{0,-1,-1,-1}
        -2 G_{0,-1,-1,0}
        +G_{0,-1,0,-1}
\big)
-32 x P_5 \big(
        2 G_{0,0,-1,0}
        -G_{0,0,1,0}
\big)
\nonumber\\
& 
+8 \big(
        17 (1+x) (3 x-1) G_{-1}
        -17 x (3 x-1) G_0
        -8 (x-3) x G_{0,-1}
        +x (11 x-8) G_{0,0}
        +(x-1) (11 x-5) G_{1,0}
\nonumber\\
& 
        -12 x^2 G_{0,-1,-1}
        -12 x (3 x-4) G_{0,0,-1}
        -6 x^2 G_{0,1,0}
        +16 (x-1) x G_{0,0,-1,-1}
        -12 x^2 G_{0,0,0,-1}
\big)
-16 x (5 x-2) \zeta_2
\nonumber\\
& 
+\big(
        32 \big(
                 6 G_{1,0,-1}
                -3 G_{1,0,0}
                -4 G_{1,0,-1,-1}
                +6 G_{1,0,0,-1}
                -G_{1,0,0,0}
        \big)
        -32 \big(
                 3 G_1
                +4 G_{1,0}
        \big) \zeta_2
        -128 \zeta_3 G_1
\big) P_2
\nonumber\\
& 
+\big(
        32 \big(
                G_{-1,0}
                +6 G_{-1,-1,-1}
                -6 G_{-1,-1,0}
                +3 G_{-1,0,-1}
        \big)
        -144 G_{-1,-1}
\big) P_3
\nonumber\\
& 
+\big(
        (16 x) \big(
                4 G_{0,-1,0}
                +G_{0,0,0}
        \big)
        +64 x \zeta_3
        +64 x \zeta_2 G_0
\big) P_4
 \Big\} \,.
\end{align}
where
$P_1 = 1-4 x+x^2, P_2 = (1-x)^2, P_3 = (x^2-1), P_4 = 3 x - 2, P_5 = x-2$.
$G_{a_1,\ldots,a_n}$ denotes the generalized poly-logarithms with argument $x = -m_Z^2/m_W^2$.
Below, we present the \oaasas partonic cross-sections for inclusive $Z$-boson production 
in the threshold limit, evaluated at the central scale choice $\mu_R^2=\mu_F^2=m_Z^2$.
Numerical values are obtained using the input parameters defined above.
\begin{align}
\Delta_{q,{\rm sv}}^{(2,1),\gamma} &=
\frac{(v_u^2+a_u^2)}{c_w^2 s_w^2} \bigg[
3 C_F^2 e_u^2 \bigg\{
\delta_z  \Big(
        -\frac{5599}{6}
        -\frac{130}{3} \zeta_2
        +\frac{412}{5} \zeta_2^2
        -\frac{184736}{315} \zeta_2^3
        -460 \zeta_3
        +80 \zeta_2 \zeta_3
        +\frac{10336}{3} \zeta_3^2
        +1328 \zeta_5
\Big)
\nonumber\\&
+ \big(
        -4096 \zeta_3
        -6144 \zeta_2 \zeta_3
        +12288 \zeta_5
\big) \cD_0
+\big(
        2044
        +2976 \zeta_2
        -\frac{14208}{5} \zeta_2^2
        -960 \zeta_3
\big) \cD_1
+10240 \zeta_3 \cD_2
\nonumber\\
&
-\big(
         2048
        +3072 \zeta_2
\big) \cD_3
+512 \cD_5
\bigg\}
+ C_A C_F e_u^2  \bigg\{
\delta_z  \Big(
        \frac{74321}{36}
        -\frac{13186}{27} \zeta_2
        -\frac{1664}{135} \zeta_2^2
        -\frac{20816}{315} \zeta_2^3
        -\frac{20156}{9} \zeta_3
\nonumber\\
&
        +\frac{28736}{9} \zeta_2 \zeta_3
        +\frac{3280}{3} \zeta_3^2
        -\frac{39304}{9} \zeta_5
\Big)
+\Big(
        \frac{25856}{27}
        -\frac{12416}{27} \zeta_2
        +\frac{1408}{3} \zeta_2^2
        +\frac{26240}{9} \zeta_3
        -1472 \zeta_2 \zeta_3
\Big) \cD_0
-\Big(
         \frac{35572}{9}
\nonumber\\
&
        +\frac{11648}{9} \zeta_2
        -\frac{3648}{5} \zeta_2^2
        +5184 \zeta_3
\Big) \cD_1
- \Big(
         \frac{4480}{9}
        -\frac{11264 \zeta_2}{3}
        -1344 \zeta_3
\Big) \cD_2
+\Big(
        \frac{17152}{9}
        -512 \zeta_2
\Big) \cD_3
-\frac{7040}{9}  \cD_4
\bigg\}
\nonumber\\
&
+ C_F T_F \sum_q e_q^2 \bigg\{
\delta_z  \Big(
        \frac{3529}{9}
        -\frac{872}{9} \zeta_2
        -\frac{96}{5} \zeta_2^2
        -\frac{656}{3} \zeta_3
        +\frac{128}{3} \zeta_2 \zeta_3
\Big)
+\Big(
        \frac{6844}{27}
        -64 \zeta_2
        -\frac{128}{5} \zeta_2^2
        -\frac{1216}{9} \zeta_3
\Big) \cD_0
\nonumber\\
&
-\Big(
         \frac{880}{3}
        -256 \zeta_3
\Big) \cD_1
+64 \cD_2
\bigg\}
%
+ C_F n_F T_F e_u^2 \bigg\{
\delta_z  \Big(
        -\frac{6055}{9}
        +\frac{7880}{27} \zeta_2
        +\frac{3136}{135} \zeta_2^2
        +\frac{8992}{9} \zeta_3
        -\frac{11392}{9} \zeta_2 \zeta_3
\nonumber\\
&
        +\frac{11072}{9} \zeta_5
\Big)
+\Big(
        -\frac{7168}{27}
        +\frac{5632}{27} \zeta_2
        -\frac{512}{3} \zeta_2^2
        -\frac{10240}{9} \zeta_3
\Big) \cD_0
+\Big(
        \frac{11216}{9}
        +\frac{4096 \zeta_2}{9}
        +2304 \zeta_3
\Big) \cD_1
+\Big(
        \frac{512}{9}
\nonumber\\
&
        -\frac{4096 \zeta_2}{3}
\Big) \cD_2
-\frac{5120}{9} \cD_3
+\frac{2560}{9} \cD_4
\bigg\}
\bigg]
+ \frac{v_u e_u \sum_q (e_q v_q)}{c_w^2 s_w^2} \bigg[
C_F \bigg\{
        96
        +240 \zeta_2
        -\frac{48}{5} \zeta_2^2
        +112 \zeta_3
        -640 \zeta_5
\bigg\}
\bigg]
\nonumber\\
&
+ \frac{a_u \sum_q (a_q e_q^2)}{c_w^2 s_w^2} \bigg[ C_F T_F \left\{ 64 \zeta_2 \right\} \bigg].
%
\\
\Delta_{q,{\rm sv}}^{(2,1),Z} &=
\frac{(v_u^4+6 v_u^2 a_u^2+a_u^4)}{3 c_w^4 s_w^4} \bigg[
C_F^2 \bigg\{
\delta_z {\cal C}_{1}^{\scriptscriptstyle Z} 
+\big(
        -8448 \zeta_3
        +6144 \zeta_2 \zeta_3
\big) \cD_0
+\Big(
        9792
        -12288 \lifourH
        +192 \zeta_2
        -\frac{12288}{5} \zeta_2^2
\nonumber\\
&
        -2688 \zeta_3
        +6912 \zeta_2 \log (2)
        -6144 \zeta_2 \log^2(2)
        -512 \log^4(2)
\Big) \cD_1
-\big(
         4224
        -3072 \zeta_2
\big) \cD_3
\bigg\}
%
+ C_A C_F \bigg\{ 
\delta_z \, {\cal C}_{2}^{\scriptscriptstyle Z} 
\nonumber\\
&
+\Big(
        \frac{17776}{9}
        -\frac{30352}{9} \zeta_2
        +1408 \zeta_2^2
        -1848 \zeta_3
        +1344 \zeta_2 \zeta_3
\Big) \cD_0
+\Big(
        -\frac{11792}{3}
        +\frac{11744}{3} \zeta_2
        -768 \zeta_2^2
\Big) \cD_1
\nonumber\\
&
+\big(
        1936
        -1408 \zeta_2
\big) \cD_2
\bigg\}
+ C_F n_F T_F \bigg\{
\delta_z   \, {\cal C}_{3}^{\scriptscriptstyle Z}
+\Big(
        -\frac{4928}{9}
        +\frac{9920}{9} \zeta_2
        -512 \zeta_2^2
\Big) \cD_0
+\Big(
        \frac{3520}{3}
        -\frac{2560 \zeta_2}{3}
\Big) \cD_1
\nonumber\\
&
+\big(
        -704
        +512 \zeta_2
\big) \cD_2
\bigg\} 
\bigg] 
%
%
{
+ \frac{(v_u^2+a_u^2)\sum_f (v_f^2+a_f^2) }{c_w^4 s_w^4} \bigg[ C_F n_F T_F ~{\cal C}_{4}^{\scriptscriptstyle Z} \bigg]
+ \frac{a_u \sum_f a_f v_f^2}{c_w^4 s_w^4} \bigg[ C_F T_F {\cal C}_{5}^{\scriptscriptstyle Z} \bigg] .
}
\end{align}
The ${\cal C}_i^{\scriptscriptstyle Z}$s are given by 
\begin{align}
 {\cal C}_{1}^{\scriptscriptstyle Z} &= -764.321004478814, ~~
 {\cal C}_{2}^{\scriptscriptstyle Z} = 659.716981015678, ~~
 {\cal C}_{3}^{\scriptscriptstyle Z} = -8.13914206965976,
 \nonumber\\
 {\cal C}_{4}^{\scriptscriptstyle Z} &= -181.8731252209420, ~~
 {\cal C}_{5}^{\scriptscriptstyle Z} = 124.5366691992181.
\end{align}
\begin{align}
 \Delta_{q,{\rm sv}}^{(2,1),W} &= 
 \frac{1}{c_w^2 s_w^4} \Big[
 (v_u+a_u)(v_d+a_d) \{ 
 C_F^2 (-120.8360539829785\, \delta_z + 390.4344767767053\, \cD_0 - 551.4221954490355\, \cD_1 
\nonumber\\
& 
 + 162.4026598701161\, \cD_3)
+ C_A C_F (61.3459\, \delta_z + 131.9091920822334\, \cD_0 + 84.33928043735597\, \cD_1 
\nonumber\\
& 
- 74.43455244046987\, \cD_2)
+ C_F n_F T_F (13.8406\, \delta_z 
- 23.47141464469002\, \cD_0 - 45.11184996392113\, \cD_1 
\nonumber\\
& 
+ 27.06710997835268\, \cD_2)
\}
+ (v_u+a_u)^2 \{ 
C_F^2 (46.13824661928950\, \delta_z - 208.8335564028634\, \cD_0 
\nonumber\\
&
+ 265.2324101020284\, \cD_1 - 86.86508760689552\, \cD_3) 
+C_A C_F (6.67222\, \delta_z - 70.55490061272370\,  \cD_0 
\nonumber\\
&
- 45.11095440033213\, \cD_1 + 39.81316515316044\, \cD_2 )
+ C_F n_F T_F (-17.4438\, \delta_z + 12.55426783649582\, \cD_0 
\nonumber\\
&
+ 24.12919100191542\, \cD_1 - 14.47751460114925\, \cD_2)
\}
+ (v_u^2+a_u^2) \{ C_F n_F T_F (3.676970001925688\, \delta_z) \}
\nonumber\\
& 
+ a_u (\sum_q v_q) \{ C_F T_F (2.479977591394453 \, \delta_z) \}
\Big].
\end{align}

\end{widetext}

\end{document}